\documentclass[a4paper,11pt]{article}
\pdfoutput=1 % if your are submitting a pdflatex (i.e. if you have
\usepackage{jcappub} % for details on the use of the package, please
\usepackage[T1]{fontenc} % if needed
\usepackage{footnote}

\usepackage{graphicx,ulem}
\usepackage{longtable}
\usepackage{float}
\usepackage{dcolumn}
\usepackage{graphics,epsfig}
\usepackage{amsmath,amssymb,latexsym,mathrsfs}
\usepackage{bm}
\usepackage{color}
\usepackage{color}
\usepackage{subfigure}
\usepackage{multirow}
\usepackage{amsfonts}

\begin{document}

\title{ Testing Predictions of the Quantum Landscape Multiverse 2: The Exponential Inflationary Potential}

\author[1,2]{Eleonora Di Valentino}
\author[3]{Laura Mersini-Houghton} 

\affiliation[1]{Institut d' Astrophysique de Paris (UMR7095: CNRS \& UPMC-Sorbonne Universities), F-75014, Paris, France}
\affiliation[2]{Sorbonne Universit\'es, Institut Lagrange de Paris (ILP), F-75014, Paris, France}
\affiliation[3]{Department of Physics and Astronomy,
    UNC-Chapel Hill, NC 27599, USA}

%\Cambridge %mersini@physics.unc.edu
%\
\date{\today}

\abstract{The 2015 Planck data release tightened the region of the allowed inflationary models. Inflationary models with convex potentials have now been ruled out since they produce a large tensor to scalar ratio. Meanwhile the same data offers interesting hints on possible deviations from the standard picture of CMB perturbations. Here we revisit the predictions of the theory of the origin of the universe from the landscape multiverse for the case of exponential inflation, for two reasons: firstly to check the status of the anomalies associated with this theory, in the light of the recent Planck data; secondly, to search for a counterexample whereby new physics modifications may bring convex inflationary potentials, thought to have been ruled out, back into the region of potentials allowed by data. Using the exponential inflation as an example of convex potentials, we find that the answer to both tests is positive: modifications to the perturbation spectrum and to the Newtonian potential of the universe originating from the quantum entanglement, bring the exponential potential, back within the allowed region of current data; and, the series of anomalies previously predicted in this theory, is still in good agreement with current data. Hence our finding for this convex potential comes at the price of allowing for additional thermal relic particles, equivalently dark radiation, in the early universe. }

%\pacs{04.25.D-, 04.25.dg, 04.30.-w, 04.30.Db}
% 04.25.D- Numerical relativity
% 04.25.dg Numerical studies of black holes and black-hole binaries
% 04.25.Nx Post-Newtonian approximation; perturbation theory; related approximations
% 04.30.-w Gravitational waves (see also 04.80.Nn Gravitational wave detectors and experiments)
% 04.30.Db Wave generation and sources
% 02.70.Hm Spectral methods

\maketitle
\flushbottom

%#########################
\section{Introduction}

How were the initial conditions of the universe selected? Probing this fundamental questions is key to understanding the universe we inhabit. Fortunately, major advances in precision cosmology, notably the Planck data \cite{planck1,planck2} and SN1a data \cite{R16} recently, place us in an unusually favorable position - for the first time we can test and falsify theories that extend physics beyond inflation and the standard model of cosmology.

In previous work \cite{archillmh,richlmh,tomolmh,lmh}, we developed a theory of the initial conditions of the universe from a quantum multiverse. The selection of the initial states was derived from the quantum dynamics of gravity acting on the string landscape, which we took as our working example for the configuration space of initial states. Allowing the wavefunctional of the universe to propagate through the landscape vacua, enabled us to derive the most probable wavefunction solution by means of the quantum cosmology formalism. The wavefunction operated on a minisuperspace spanned by the scale factor $a$ of 3-geometries and the landscape moduli variables $\phi$ with the landscape vacua energy profile, the potential $V(\phi)$. The solutions we found, with the decoherence among branches included \cite{richlmh} showed why high energy inflationary states are natural.

Tracing out long wavelength modes of metric and matter fluctuations decohered the wavefunction of the universe \cite{richlmh, tomolmh}. Since coherence and decoherence are closely related, the latter induced a nonlocal entanglement among all surviving branches of the wavefunction, localized at various high energy landscape vacua, and which give rise to universes like ours. We calculated the strength of the entanglement of our branch with others for convex potentials in \cite{tomolmh}, and for concave potentials in \cite{lmh}, and the modifications it induced on the Friedmann equation and the gravitational potential of the universe. The energy of landscape vacua varies across the landscape. This energy for each vacuum is denoted by the parameter $b$, which is local since it differs from one vacuum to the next. Assuming that locally the vacuum states may acquire their energy from SUSY breaking, we herein refer to the parameter $b$ as the 'SUSY breaking parameter'. It is possible that vacua may acquire their energy $b$ from some other mechanisms instead of SUSY breaking. Therefore, $b$ should be seen as a label of the individual vacuum energies independently of the mechanism that produced them.

In this work we analyze the status with the predictions of this theory of the origin of the universe from a quantum landscape multiverse \cite{archillmh,richlmh,tomolmh,lmh}, against the most recent Planck data \cite{planck1, planck2}, for the case of a convex potential, the exponential inflation. The study of concave inflationary potentials was given in \cite{edvlmh}. The modifications to the perturbation spectrum and the Newtonian potential of the universe, originating from quantum entanglement in this theory, were derived in \cite{tomolmh} for the case of exponential inflation. We review the main results below. In its simplest and unmodified form, based on a 6 parameter model of standard cosmology, the exponential inflationary potential is already ruled out. In fact based on the 6 parameter model of the standard cosmology, Planck's recent bounds on the tensor to scalar ratio $r$ rule out most of the convex potentials, since they predict large $r$'s. 

The aim of our study here is two fold: firstly we aim to compare the predictions of the theory of the origin of the universe from the quantum multiverse against the recent measurements made by Planck for convex potential such as the exponential model; secondly, we wish to demonstrate that if the standard model of cosmology were extended to allow for new physics then convex inflationary potentials that in the simple picture were thought to be ruled out by data, may not in fact be ruled out in the extended picture. The analysis of concave downard potentials, specifically the Starobinsky model, was done in \cite{edvlmh}. The point of choosing the exponential inflation as the object of our study of convex potentials, is not to advocate that this is how our universe started. Rather, we hope that by providing a counterexample to the standard interpretation of observations, in our case the chosen example is the exponential potential, may help to emphasize a key point, namely: data and our understanding of observational results are two different things. That is, ruling out models of inflation is not straightforward if we allow the 6 parameter standard model of cosmology to be extended to new physics realms. 

The paper is organized as follows: in Section \ref{sec:model} we first review the main results of the modified spectra, field solutions and Friedmann equation for the exponential potential of \cite{tomolmh}, in the context of the theory of the origin of the universe from the quantum landascape, since we need these expressions for the data analysis. In Section \ref{method} we present the method of analyzing the predictions of this theory against data, provide the results in Section \ref{results}, and conclude in Section \ref{sec:conclusions}.

\section{Convex Potentials}
\label{sec:model}

In contrast to concave downward inflationary potentials, studied in \cite{lmh, edvlmh} for the case of Starobinsky potential, the exponential inflationary model is a convex potential with $V''>0$ . In convex inflationary models like the exponential model, the mass term which drives fluctuations is $V'' = m^2 >0$. Fluctuations are small and stable in these potentials when expanded around the minimum. The correction to the Friedmann equation derived in \cite{tomolmh} for convex potentials, is: $V_{eff} =  = V - \frac{1}{2} \frac{V^2}{9M^4} F[b,V] = V + \frac{1}{2} \frac{V^2}{9M^4} |F[b,V]|$  in Eq. \ref{effectivev}, since $F[b,V] <0$ and $m^2 = V" >0$.

Einstein equations are modified accordingly since the inflaton potential $V$ is now replaced by the modified inflationary potential $V_{eff}$, which besides its dependence on the field $\phi$, also depends on the landscape 'Susy breaking parameter' $b$ . $V_{eff}$ encaptures the nonlocal entanglement between our wavefunction branch and other branches.
The energy correction term in $V_{eff}$

\begin{equation}
V_eff = V - f[b,V] = V + | f[b,V] |
\label{effectivev}
\end{equation}

is denoted by $f[b,V]$ where

\begin{equation}
f(\phi)=\frac{1}{2} \left[ \frac{V(\phi)}{3M_P^2} \right]^2 F(\phi)
\label{energycorrection}
\end{equation}

with $b$ the landscape parameter describing the energy scale from 'SUSY breaking' in each vacua. The entanglement information contained in $F[b,V]$ was derived in \cite{tomolmh}, and is given by

\begin{equation}
F(\phi) = \frac{3}{2}\left(2+\frac{m^2M_P^2}{V(\phi)}\right)Log\left(\frac{b^2M_P^2}{V(\phi)}\right)-\frac{1}{2}\left(1+\frac{m^2}{b^2}\right)e^{-\frac{3b^2M_P^2}{V(\phi)}}
\end{equation}

The Friedman equation of expansion becomes

\begin{equation}
3 M_{p}^2 H^2 = V_{eff} = V - f[b,V]
\label{hubble}
\end{equation}

for convex potentials. We apply these modifications to the exponential potential as an example of applying the quantum entanglement modifications in the theory of the origin of the universe from a quantum landscape to the convex types.
The field solution is modified accordingly since the inflationary potential $V(\phi)$ is replaced with the modified potential $V_{eff} (\phi, b)$. The field solution is obtained from the field equation

\begin{equation}
3 H d\phi /dt = -\frac{\partial V_{eff}}{\partial d\phi}
\label{fieldsol}
\end{equation}

Using $d\phi/dt = H d\phi/dlnk$,the field equation Eq. \ref{fieldsol}, from Eq.\ref{hubble}, becomes

\begin{equation}
dN = \frac{V_{eff}}{M_{p}^2 V'_{eff}} d \phi = - d lnk
\label{fieldwithk}
\end{equation}

We assume the correction term is small compared to the leading term $V$ i.e. slow roll holds even with correction terms. Then we can approximate the integral in Eq.\ref{fieldwithk} as: $\int \frac{V_eff}{V'_{eff}} d\phi \simeq \left(\frac{1- f/V}{1 - df/dV}\right)\int \frac{V}{V'}$.

This last equation \ref{fieldwithk} gives us the field as a function of $k$, or equivalently the number of efolds $dN$. Integrating $dN$ from the start to the end of slow roll, allows us to get the total number of efolds $N$.

%%%%%%%%%%%%%%%%%%%%%%%%%%%%%%%%%%%%%%%%
%%%%%%%%%%%%%%%%%%%%%%%%%%%%%%%%%%%%%%%%
\subsection{Exponential Potential} 

Let us now apply these modifications to the exponential potential as an example of the modified convex potentials reviewed above. The derivation of modifications to the perturbation spectrum and the Newtonian potential of the universe, and the series of predictions resulting from these modifications were derived in \cite{tomolmh}. However the only data available then was the first release of WMAP data. One of our goals here is to check the status of the predictions made in \cite{tomolmh}, and specifically the status of anomalies we predicted there over a decade ago, in the light of the most recent Planck data release \cite{planck1, planck2} and the new release of the SN1a data of Riess et al. \cite{R16}.

The pure unmodified exponential potential, introduced the first time by \cite{Lucchin:1984yf}, is

\begin{equation}\label{exp_pot}
V(\phi) = V_0 e^{-\frac{\lambda\phi}{M_P}}
\end{equation}

and $V'=dV/d\phi$ is

\begin{equation}
V'(\phi) = -\frac{V_0\lambda}{M_P} e^{-\frac{\lambda\phi}{M_P}}
\end{equation}

The curvature of the potential is positive and the mass squared term which drives the fluctuations, is given by the curvature $V''$ evaluated at the start of slow roll. So far, the expression for the effective potential and the correction term $f(b, V)$ are provided with respect to $M_p$ which denotes the normal Planck mass.

The mass term that drives the fluctuations is

\begin{equation}
m^2= V''(\phi_{in}) = V_{0}\left(\frac{\lambda}{M_P}\right)^{2} e^{-\frac{\lambda\phi_{in}}{M_P}} \approx \frac{V_{0} \lambda^2}{M^2}
\end{equation}

The quantities we need for carrying out the data analysis in the next section, such as the power spectrum, field solution, the tensor and scalar index, are calculated \cite{tomolmh} from the modified potential which includes the correction term $V_{eff}$. The derivative of the effective potential is 

\begin{equation}
V'_{eff}(\phi)=V'(\phi)(1  - df(\phi)/dV)
\end{equation}

where 

\begin{equation}
\frac{df}{dV}= \frac{V}{9 M_{p}^4} \left(F[b,V] + \frac{V}{2} dF/dV \right)
\end{equation}

and
\begin{equation}
\frac{dF}{dV} = -\frac{3\left(m^2 M_{p}^2 -M^2 (b^2 +m^2) Exp[-\frac{3b^2 M_{p}^2}{V}] -2V - m^2 M_{p}^2 Log[\frac{b^2 M_{p}^2}{V}]\right)}{2 V^2}
\label{dFdv}
\end{equation}

The second derivative of the effective potential needed below is
\begin{equation}
V''_{eff}(\phi)=V''(\phi)\left(1 - \frac{df}{dV}\right) -\frac{d^2f}{dV^2} V'^2
\end{equation}

where $V'' = (\lambda / M_{p})^2 V$.
The unmodified field solution for the exponential potential obtained from integrating the field equation, gives

\begin{equation}
\phi_0(k)=\lambda M_P ln \left( \frac{k}{k_*} \right)
\end{equation}

In the presence of modifications, with the potential $V_{eff}$, the field satisfies the equation

\begin{equation}
\frac{V_{eff}}{M_{p}^2 V'_{eff}} d\phi = - dLn(k)
\end{equation}

which is integrated to give the modified field solution

\begin{equation}
\phi(k)=\lambda M_P ln \left( \frac{k}{k_*} \right)\left[ \frac{V'_{eff}(\phi_0)}{V'(\phi_0)} \right]\left[ \frac{V(\phi_0)}{V_{eff}(\phi_0)} \right] = \phi_{0}(k) \frac{(1 -df/dV)}{(1 - f/V)}
\end{equation}

We assumed that slow roll holds in the presence of modification, i.e the condition $\frac{\Delta V_{eff}}{\Delta \phi^4} \ll 10^{-7}$ holds. Therefore variations in the correction term $f(b, V)$ are bound to be less than $7$ orders of magnitude during the slow roll regime, which allows us to pull them out of the field integral and approximate $f(b, V)$ and $f'(b, V)$ with their value at the start of slow roll.

We can now put everything together to calculate the power spectrum and tensor to scalar ratio. Note that the expressions of the spectra are defined with respect to the reduced Planck mass, obtained from the Planck mass by a rescaling of $M_p$ with: $M_{p} / \sqrt{8 \pi}$). Since in what follows our data analysis and spectra are estimated with respect to the reduced Planck mass, then we can continue using the reduced Planck mass and equivalently rescale $b$ by the factor of $\sqrt{8 \pi}$ because $b$ always enters in a ratio $b/M_p$ in the above modifications.

\begin{equation}
P_\zeta(k)=\frac{1}{24\pi^2M_P^6}\left[ \frac{V_{eff}(\phi)^3}{V'_{eff}(\phi)^2} \right] \simeq \frac{1}{24\pi^2M_P^6}\left[ \frac{V(\phi)^3}{V'(\phi)^2} \right] \frac{(1 - f/V)^3}{(1-df/dV)^2}
\end{equation}

All unmodified quantities are denoted by $_0$, eg $\phi_0 , r_0 , P_0$ etc, mean unmodified and they are evaluated with respect to the unmodified field $\phi_{0}[k]$. All modified quantites are evaluated with respect to the modified field solution $\phi[k]$.

 The modifed tensor to scalar ratio is 
\begin{equation} 
r [k] = 8 M_{p}^2 \left(\frac{V'_{eff}(\phi)}{V_{eff}(\phi)}\right)^{2} 
\end{equation}

To calculate $P[k], r[k], n[k]$ we only need $V'$ and $V'_{eff}$ which we calculated above. With the above reminder that the modified field solution must be used to estimate the modified quantities, we can approximately relate $r[k]$ to $r_{0}[k]$ by $r[k] \approx r_{0}[k]  (\frac{( 1 - df/dV )}{( 1 - f/V )})^2$, bearing in mind that in the approximate relation of $r[k]$ with $r_{0}[k]$ both are evaluated with respect to $\phi$, not $\phi_{0}$.

From the modified power spectrum we can then calculate the modified scalar index $n[k] - 1 = dLog(P[k])/dLog(k)$ with $P[k]$ given above. The unmodified scalar tensor is $n_{0}[k] -1 =  dLog(P_0[k])/dLog(k)$ where the $_0$ notation means unmodified field and power spectrum and scalar tensor.

\section{Method} \label{method}
We explore the modified Exponential model, by considering the $8$ cosmological parameters listed in the following, that are explored within the range of the conservative flat priors reported in Table~\ref{priors}. We vary the $4$ standard parameters of the $\Lambda$CDM, namely: the baryon $\Omega_bh^2$ and the cold dark matter energy densities $\Omega_ch^2$, the reionization optical depth $\tau$ and the ratio between the sound horizon and the angular diameter distance at decoupling $\Theta_{s}$. Moreover, we consider $3$ inflationary parameters: the effective inflationary potential parameter $\lambda$ (see eq.~\ref{exp_pot}), the energy scale of the inflation $10^{10}V_0/M^4$, and the contribution of the primordial gravitational waves with a tensor-to-scalar ratio of amplitude $r$ at the pivot scale $k_0=0.002 hMpc^{-1}$. Finally, we have the natural logarithm $log(\sqrt{8\pi}b[GeV])$ of the SUSY-breaking scale associated with the landscape effects $b$.
As a second step, we consider extensions to these baseline model, by adding the effective number of relativistic degrees of freedom $N_{\rm eff}$.

\begin{figure}
\centering
\includegraphics[width=7.4cm]{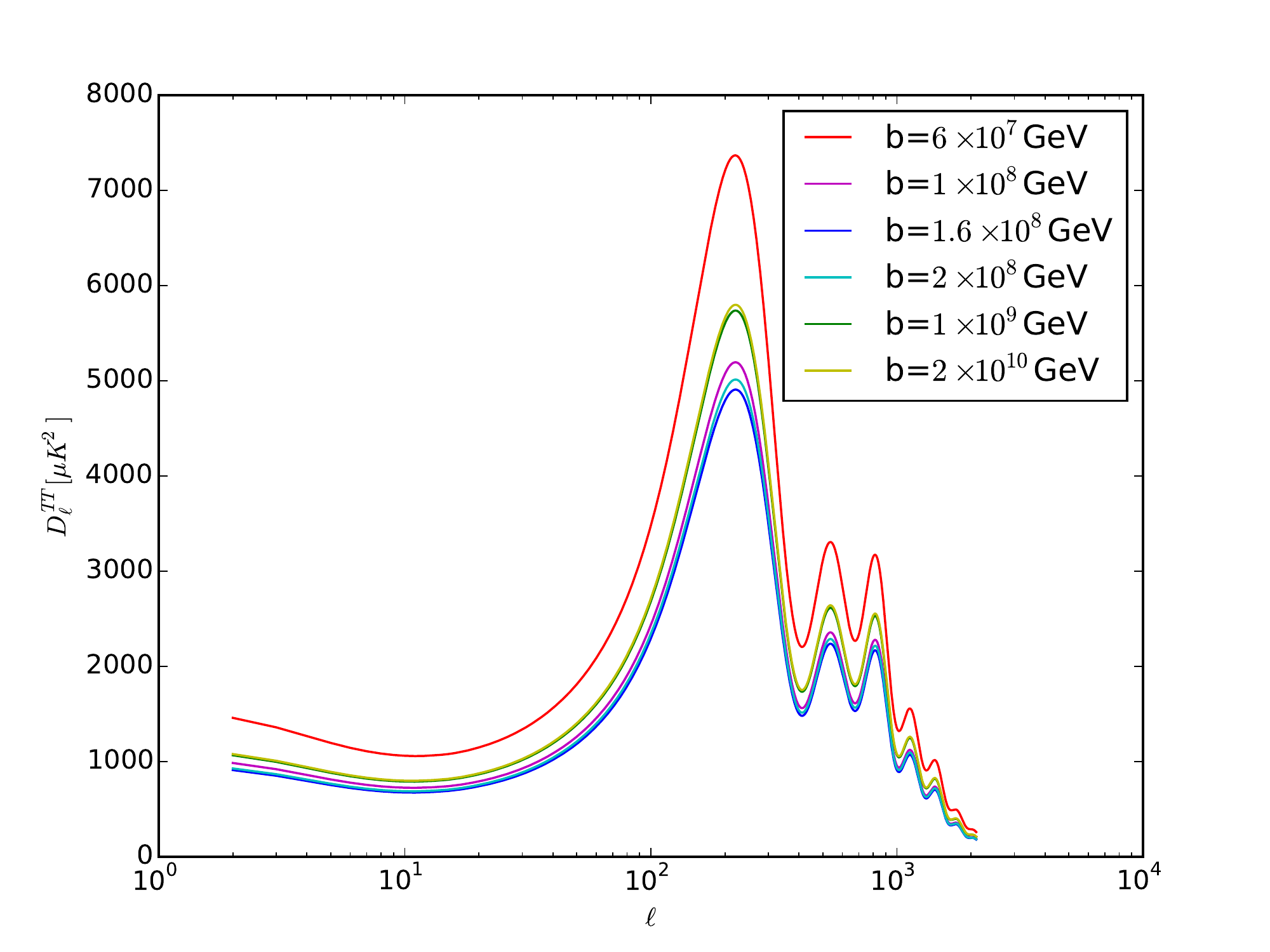}
\includegraphics[width=7.4cm]{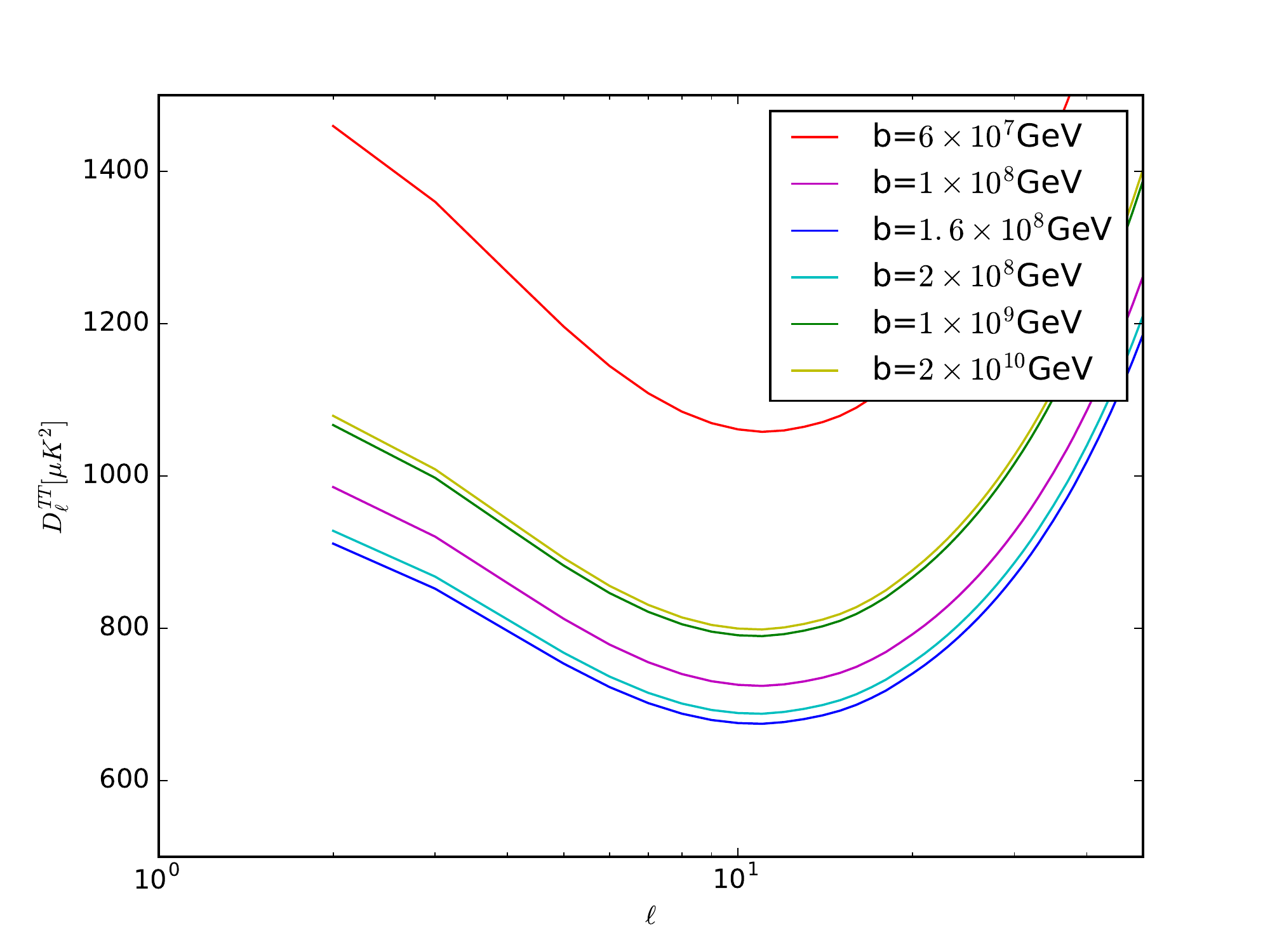}
\includegraphics[width=7.4cm]{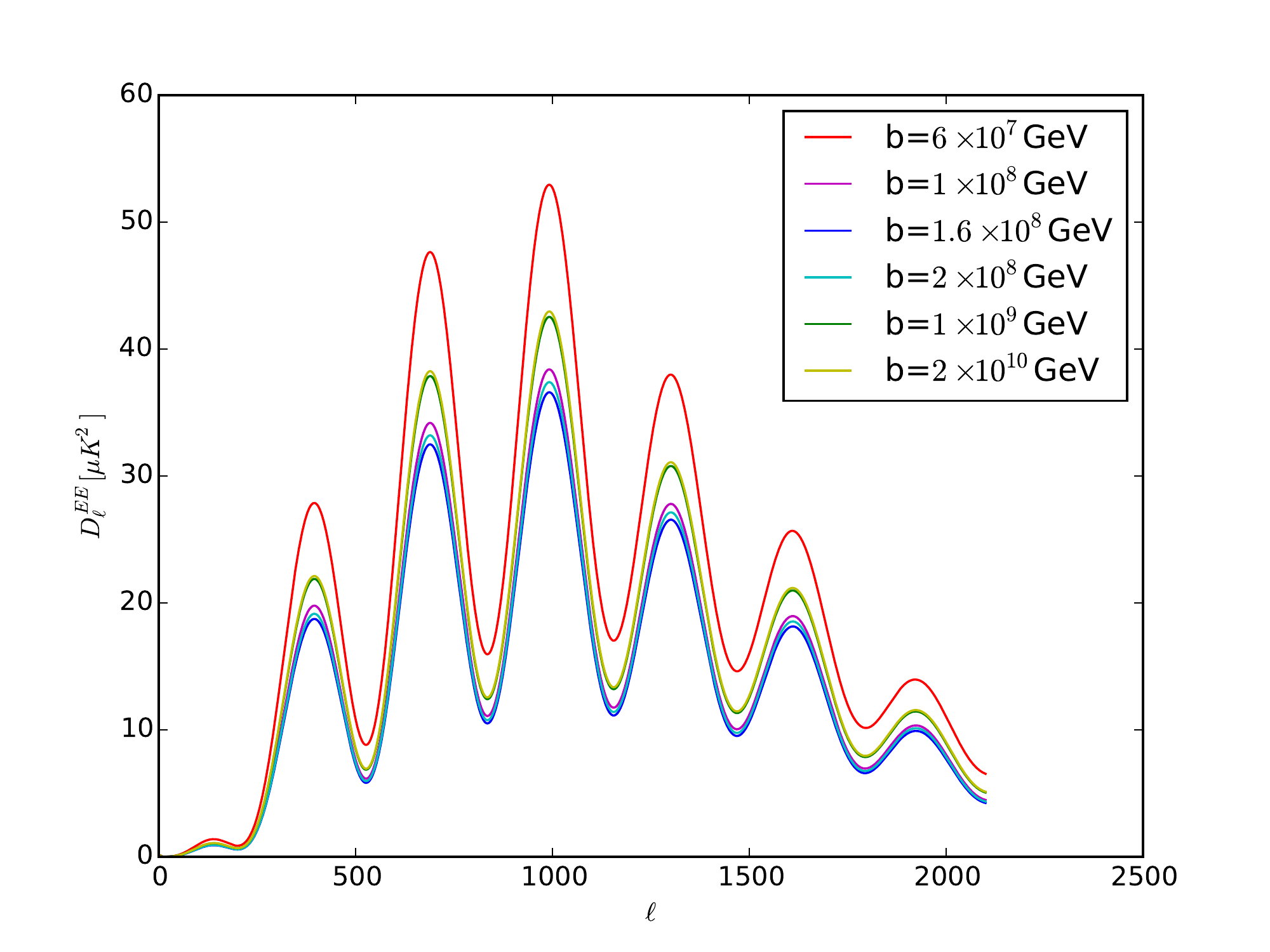}
\includegraphics[width=7.4cm]{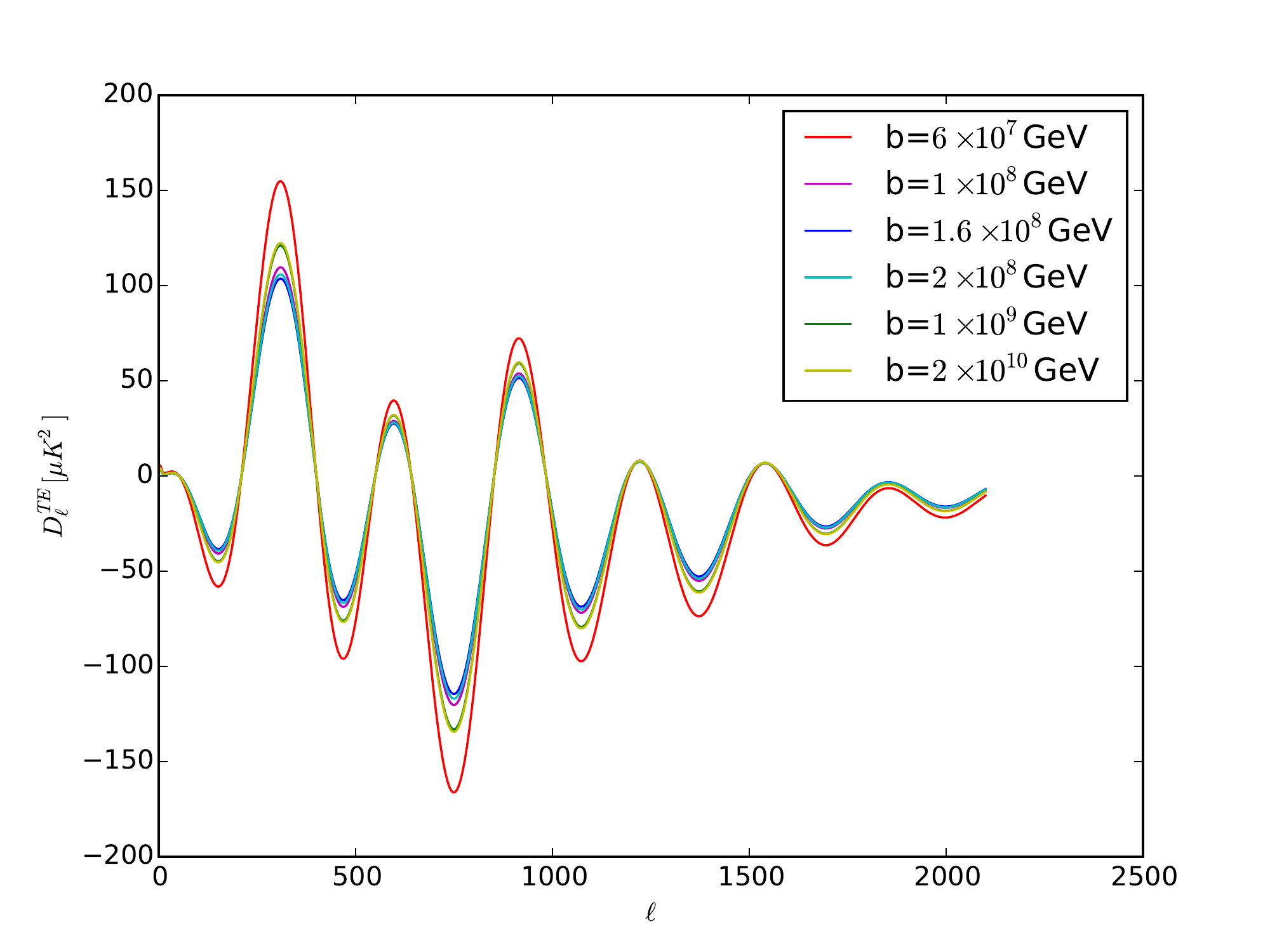}
\caption{Temperature and polarization CMB angular power spectra by varying the SUSY-breaking scale $b$ associated with the landscape effects, for the modified Exponential model.}
\label{clparamE}
\end{figure}

In Fig.~\ref{clparamE} we show the temperature and polarization CMB angular power spectra by varying the SUSY-breaking scale $b$ associated with the landscape effects.

\begin{table}
\begin{center}
\begin{tabular}{|c|c|c|}
\hline
Parameter                   & Exponential\\
\hline
$\Omega_{\rm b} h^2$          & $[0.005,0.1]$\\
$\Omega_{\rm cdm} h^2$      & $[0.001,0.99]$\\
$\Theta_{\rm s}$             & $[0.5,10]$\\
$\tau$                       & $[0.01,0.8]$\\
$log(\sqrt{8\pi}b[GeV])$         & $[15,25]$\\
$10^{10}V_0/M^4$     & $[20,80]$\\
$\lambda$                 & $[0.05,0.15]$\\
$r$ &  $[0,3]$\\
$N_{\rm eff}$ & $ [0.05,10]$\\
\hline
\end{tabular}
\end{center}
\caption{External priors on the cosmological parameters assumed in this work.}
\label{priors}
\end{table}

We analyzed our model by considering several current cosmological probes.

Our baseline dataset consists of the ``Planck TT + lowTEB'' data provided by the Planck collaboration \cite{Aghanim:2015xee}. This dataset includes the full range of the 2015 temperature power spectrum ($2\leq\ell\leq2500$) in combination with the low-$\ell$ polarization power spectra in the multipoles range $2\leq\ell\leq29$.
As a second step, we use the ``Planck TTTEEE + lowTEB'' data, where we add the high multipoles Planck polarization spectra \cite{Aghanim:2015xee}, in the range $30\leq\ell\leq2500$, by taking into account that this combination of datasets is considered less robust than ``Planck TT + lowTEB'' by the Planck collaboration.
In order to check the robustness of our results, we also replace the lowTEB data with a gaussian prior on the reionization optical depth $\tau=0.055\pm0.009$, that we call ``tau055'' as obtained from the Planck HFI measurements in \cite{newtau}.
Afterwards, we consider the ``lensing'' dataset, i.e. the 2015 Planck CMB lensing reconstruction power spectrum $C^{\phi\phi}_\ell$ \cite{Ade:2015zua}.
Moreover, we use the "BAO" measurements. These include the baryonic acoustic oscillation data from 6dFGS \cite{beutler2011}, SDSS-MGS \cite{ross2014}, BOSSLOWZ \cite{anderson2014} and CMASS-DR11 \cite{anderson2014} surveys as it has been done in \cite{planckparams2015}.
Additionally, we consider the CMB polarization $B$ modes constraints provided by the 2014 common analysis of Planck, BICEP2 and Keck Array \cite{BKP}, that we call ``BKP'' dataset.
Finally, by quoting the value provided with direct measurements from SN1a in Riess et al. \cite{R16}, we include a gaussian prior on the Hubble constant $H_0=73.2\pm1.7$ km/s/Mpc, that we call "H073p2".

In order to analyze statistically our model using these datasets, to include the corrected Exponential model, we have modified the June 2016 version of the publicly available Monte-Carlo Markov Chain package \texttt{cosmomc} \cite{Lewis:2002ah}, with a convergence diagnostic based on the Gelman and Rubin statistic. As the original code, this version implements an efficient sampling of the posterior distribution using the fast/slow parameter decorrelations \cite{Lewis:2013hha}, and it includes the support for the Planck data release 2015 Likelihood Code \cite{Aghanim:2015xee} (see \url{http://cosmologist.info/cosmomc/}).

\section{Results: Modified Exponential Inflation}\label{results}

The results of our explorations are shown in Tables~\ref{table1} and \ref{table2}, where we report the constraints at $68 \% $ c.l. on the cosmological parameters. All the bounds that we will quote hereinafter are at $68\%$ c.l., unless otherwise expressed. These Tables differ for the cosmological scenario explored, respectively the $\Lambda$CDM+r and $\Lambda$CDM+r+$N_{\rm eff}$.

\begin{table*}
\begin{center}\footnotesize
\scalebox{0.9}{\begin{tabular}{lccccc}
\hline \hline
      &Planck TT    &Planck TT &Planck TTTEEE &Planck TTTEEE\\                     
                   & + lowTEB     &      + lowTEB + BAO &+ lowTEB    &+ lowTEB + lensing  \\   
\hline
\hspace{1mm}\\

$\Omega_{\textrm{b}}h^2$& $0.02254\,\pm 0.00021 $& $0.02243\,\pm0.00019$ & $0.02244\,\pm 0.00014 $ & $0.02245\,^{+0.00014}_{-0.00016} $   \\
\hspace{1mm}\\

$\Omega_{\textrm{c}}h^2$& $0.1150\, \pm0.0015$& $0.1167\,\pm 0.0010$ & $0.1165\, \pm0.0010$   & $0.1163\,\pm0.0010 $  \\
\hspace{1mm}\\

$\tau$& $0.094\,\pm0.018$& $0.088\,\pm0.017$ & $0.094\,\pm0.016$  & $0.083\,\pm 0.011$ \\
\hspace{1mm}\\

$10^{10}V_0/M^4$& $101\,^{+20}_{-10}$& $114\,^{+10}_{-8}$ & $116\,^{+10}_{-7}$& $113\,^{+10}_{-6} $ \\
\hspace{1mm}\\

$log(\sqrt{8\pi}b[GeV])$& $>22.8$& $>22.8$   & $23.4\,^{+1.3}_{-0.8}$  & $>22.9$ \\
\hspace{1mm}\\

$\lambda$ &  $>0.128$ &  $>0.140$ &  $>0.140$   & $>0.140$\\
\hspace{1mm}\\

$r$ &  $0.145\,_{-0.018}^{+0.035}$ &  $0.164\,_{-0.013}^{+0.019}$  &  $0.164\,_{-0.018}^{+0.010}$   & $0.164\,_{-0.009}^{+0.019}$ \\
\hspace{1mm}\\

$H_0$ &      $67.47\,^{+0.65}_{-0.79}$&      $ 68.68\pm0.47$  &      $68.72\pm0.48$ & $ 68.83\,\pm0.48$ \\
\hspace{1mm}\\

$\sigma_8$   & $ 0.826\,\pm0.015$   & $ 0.828\,\pm0.014$ & $ 0.832\,\pm0.013$ & $ 0.8214\,\pm 0.0083$ \\
\hspace{1mm}\\

\hspace{1mm}\\
\hline
\hline

\end{tabular}}
\caption{$68 \% $ c.l. constraints on cosmological parameters in our extended $\Lambda$CDM+r scenario from different combinations of datasets with a Exponential inflation.}
\label{table1}
\end{center}
\end{table*}

Regarding the results of the Table~\ref{table1}, we note that with respect to the minimal standard cosmological model $\Lambda$CDM+r, with only 6 parameters, this modified version of the Exponential model does not improve the situation, in fact it may even provide a worse fit of the data, increasing $\Delta \bar \chi^2=9$ when considering Planck TT+lowTEB, and increasing $\Delta \bar \chi^2=13$, when considering Planck TTTEEE+lowTEB data.
As it is well known the Exponential model is excluded from the current data since predicts a large $r$. We find that with the modifications $f[b,V]$ included, the parameters $b, V_{0},\lambda$ 
preferred by data are such that the modification term to $V_{eff}$  becomes negligible. Therefore within the 6 parameter model of cosmology, the exponential model of inflation, is forced to converge towards its unmodified version, which is ruled out by data. Including our modifications to the Exponential model do not help bring it within the allowed region of best fit models.

\begin{table*}
\begin{center}\footnotesize
\scalebox{0.7}{\begin{tabular}{lccccccc}
\hline \hline
      Planck TT    & & && &&\\                     
                 & + lowTEB    &      + lowTEB + BAO  &  + lowTEB + lensing      & + lowTEB + BKP & +tau055 & + lowTEB + H073p2\\  
\hline
\hspace{1mm}\\

$\Omega_{\textrm{b}}h^2$& $0.02271\,\pm 0.00022 $& $0.02260\,\pm0.00021$    & $0.02269\,\pm0.00021 $& $0.02276\,\pm 0.00020$& $0.02250\,\pm 0.00020$ & $0.02277\,\pm 0.00021$   \\
\hspace{1mm}\\

$\Omega_{\textrm{c}}h^2$& $0.1237\, \pm0.0034$& $0.1258\,\pm 0.0033$    & $0.1222\,\pm0.0033 $& $0.1240\,\pm0.0034$ & $0.1240\,\pm0.0035$ & $0.1238\,\pm 0.0034$   \\
\hspace{1mm}\\

$\tau$& $0.092\,\pm0.019$& $0.087\,\pm0.019$    & $0.083\,\pm 0.015$& $0.098\,\pm 0.018$& $0.0617\,\pm 0.0087$ & $0.094\,\pm 0.019$    \\
\hspace{1mm}\\

$10^{10}V_0/M^4$& $57\, ^{+10}_{-30}$& $70\pm30$    & $59\,^{+10}_{-40} $& $42\,^{+7}_{-20}$& $97\,^{+28}_{-15}$ & $<56.2$   \\
\hspace{1mm}\\

$log(\sqrt{8\pi}b[GeV])$& $22\, ^{+3}_{-1}$& $>21.9$    & $>21.5$& $>21.4$& $22.4\,^{+2.2}_{-1.1}$ & $>21.2$    \\
\hspace{1mm}\\

$\lambda$ &  $0.098\,_{-0.026}^{+0.019}$ &  $0.109\,_{-0.018}^{+0.028}$  & $0.100\,_{-0.029}^{+0.019}$&  $0.085\,_{-0.020}^{+0.012}$ &  $>0.126$& $0.091\,_{-0.026}^{+0.015}$   \\
\hspace{1mm}\\

$r$ &  $0.086\,_{-0.049}^{+0.024}$ &  $0.105\pm0.039$  & $0.091\,_{-0.052}^{+0.025}$&  $0.062\,_{-0.029}^{+0.014}$ &  $0.162\,_{-0.048}^{+0.030}$& $0.074\,_{-0.043}^{+0.018}$   \\
\hspace{1mm}\\

$N_{\rm eff}$ &  $3.51\pm0.15$&      $ 3.53\pm0.17$ & $ 3.46\,\pm0.16$   &  $ 3.56\,\pm 0.14$ &  $ 3.43\,^{+0.15}_{-0.17}$  &  $ 3.55\,^{+0.16}_{-0.15}$ \\
\hspace{1mm}\\

$H_0$ &      $71.6\, ^{+1.1}_{-0.9}$&      $ 70.88\pm0.97$ & $ 71.6\,^{+1.2}_{-1.0}$   &  $ 72.12\,^{+0.88}_{-0.79}$ &  $ 70.4\,^{+0.8}_{-1.1}$  &  $ 72.1\,^{+1.0}_{-0.8}$ \\
\hspace{1mm}\\

$\sigma_8$   & $ 0.850\,\pm0.018$   & $ 0.854\,^{+0.018}_{-0.021}$   & $ 0.838\,\pm 0.011$ &  $ 0.856\,\pm0.017$ &  $ 0.827\,_{0.012}^{+0.013}$ &  $ 0.853\,^{+0.17}_{-0.19}$  \\
\hspace{1mm}\\

\hspace{1mm}\\
\hline
\hline
      Planck TTTEEE    & & && &&\\                     
          &  + lowTEB     &       + lowTEB + BAO & + lowTEB + lensing     & + lowTEB + BKP & +tau055 & + lowTEB + H073p2\\   
\hline
\hspace{1mm}\\

$\Omega_{\textrm{b}}h^2$& $0.02264\,\pm 0.00017 $& $0.02257\,\pm0.00016$    & $0.02261\,\pm0.00017 $& $0.02275\,\pm 0.00016$& $0.02251\,\pm 0.00015$ & $0.02274\,\pm 0.00017$   \\
\hspace{1mm}\\

$\Omega_{\textrm{c}}h^2$& $0.1233\, \pm0.0027$& $0.1238\,\pm 0.0027$    & $0.1220\,\pm0.0026 $& $0.1244\,\pm0.0028$ & $0.1235\,\pm0.0026$ & $0.1242\,\pm 0.0027$   \\
\hspace{1mm}\\

$\tau$& $0.092\,\pm0.017$& $0.089\,\pm0.017$    & $0.078\,\pm 0.013$& $0.100\,^{+0.017}_{-0.016}$& $0.0645\,\pm 0.0084$ & $0.096\,\pm 0.017$    \\
\hspace{1mm}\\

$10^{10}V_0/M^4$& $83\,^{+30}_{-20}$& $92\,^{+30}_{-20}$    & $85\,^{+30}_{-20} $& $55\,\pm20$& $109\,^{+20}_{-10}$ & $67\,^{+20}_{-30}$   \\
\hspace{1mm}\\

$log(\sqrt{8\pi}b[GeV])$& $>22.3$& $>22.3$    & $22.9\, ^{+2.0}_{-0.7}$& $>21.7$& $>22.3$ & $>21.9$    \\
\hspace{1mm}\\

$\lambda$ &  $0.119\,_{-0.014}^{+0.024}$ &  $>0.119$  & $0.122\,_{-0.011}^{+0.024}$&  $0.096\pm0.018$ &  $>0.137$& $0.106\,_{-0.019}^{+0.023}$   \\
\hspace{1mm}\\

$r$ &  $0.121\pm0.036$ &  $0.134\,_{-0.030}^{+0.042}$  & $0.127\,_{-0.033}^{+0.040}$&  $0.080\,_{-0.034}^{+0.023}$ &  $0.173\,_{-0.034}^{+0.018}$& $0.097\,_{-0.041}^{+0.033}$   \\
\hspace{1mm}\\

$N_{\rm eff}$ &  $3.41\pm0.13$&      $ 3.40\pm0.14$ & $ 3.36\,\pm0.13$   &  $ 3.51\,\pm 0.14$ &  $ 3.36\,\pm 0.11$  &  $ 3.49\,\pm 0.13$ \\
\hspace{1mm}\\

$H_0$ &      $70.4\,^{+0.9}_{-1.0}$&      $ 70.07\,^{+0.81}_{-0.96}$ & $ 70.4\,^{+0.9}_{-1.0}$   &  $ 71.27\,^{+0.88}_{-0.77}$ &  $ 69.67\,^{+0.60}_{-0.74}$  &  $ 71.15\,^{+0.92}_{-0.83}$ \\
\hspace{1mm}\\

$\sigma_8$   & $ 0.851\,\pm0.016$   & $ 0.850\,\pm0.016$   & $ 0.835\,\pm 0.010$ &  $ 0.861\,\pm0.015$ &  $ 0.829\,\pm0.010$ &  $ 0.856\,\pm0.016$  \\
\hspace{1mm}\\

\hspace{1mm}\\
\hline
\hline

\end{tabular}}
\caption{$68 \% $ c.l. constraints on cosmological parameters in our extended $\Lambda$CDM+r+$N_{\rm eff}$ scenario from different combinations of datasets with a Exponential inflation.}
\label{table2}
\end{center}
\end{table*}

\begin{figure}
\centering
\includegraphics[width=7.4cm]{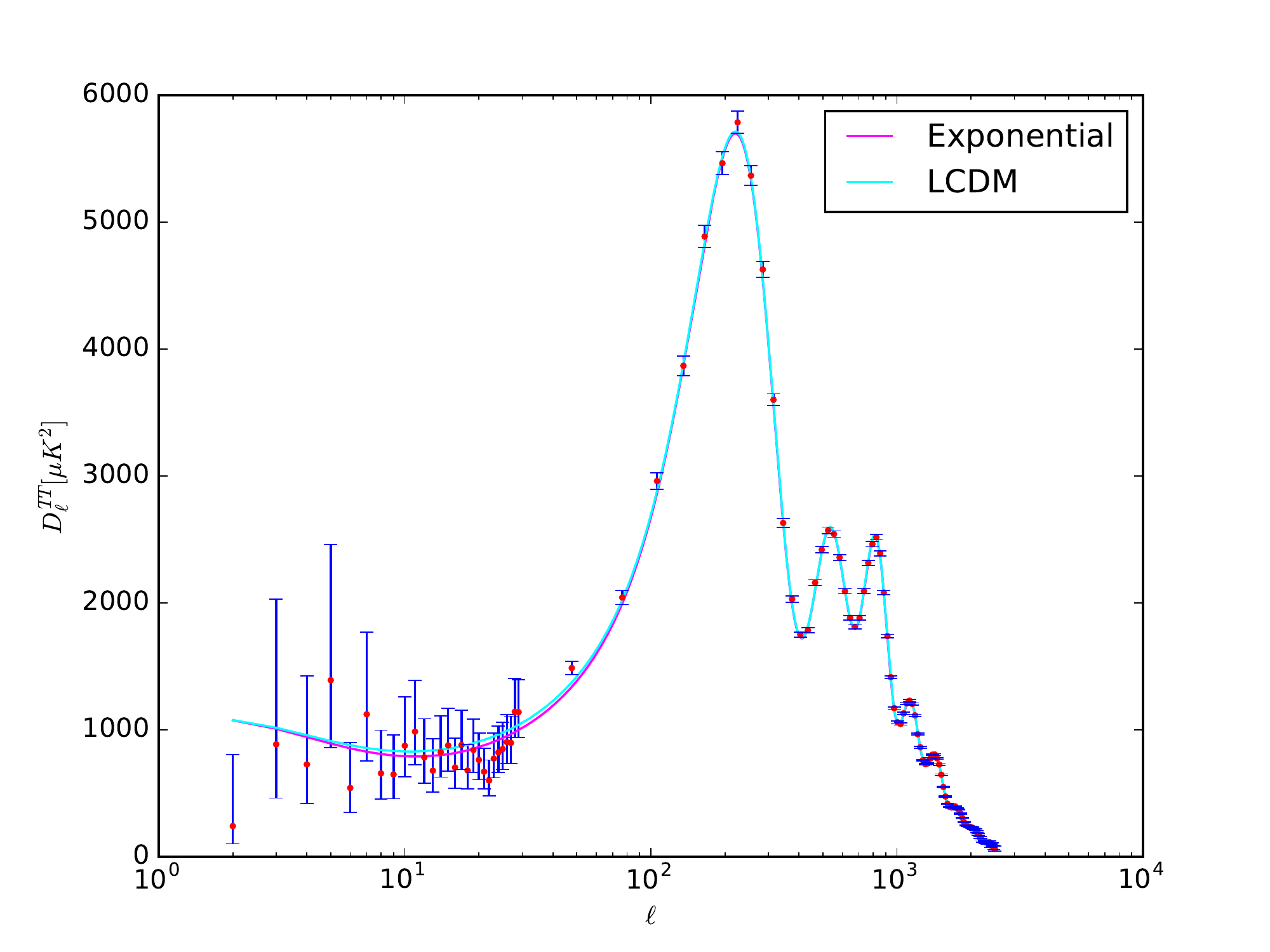}
\includegraphics[width=7.4cm]{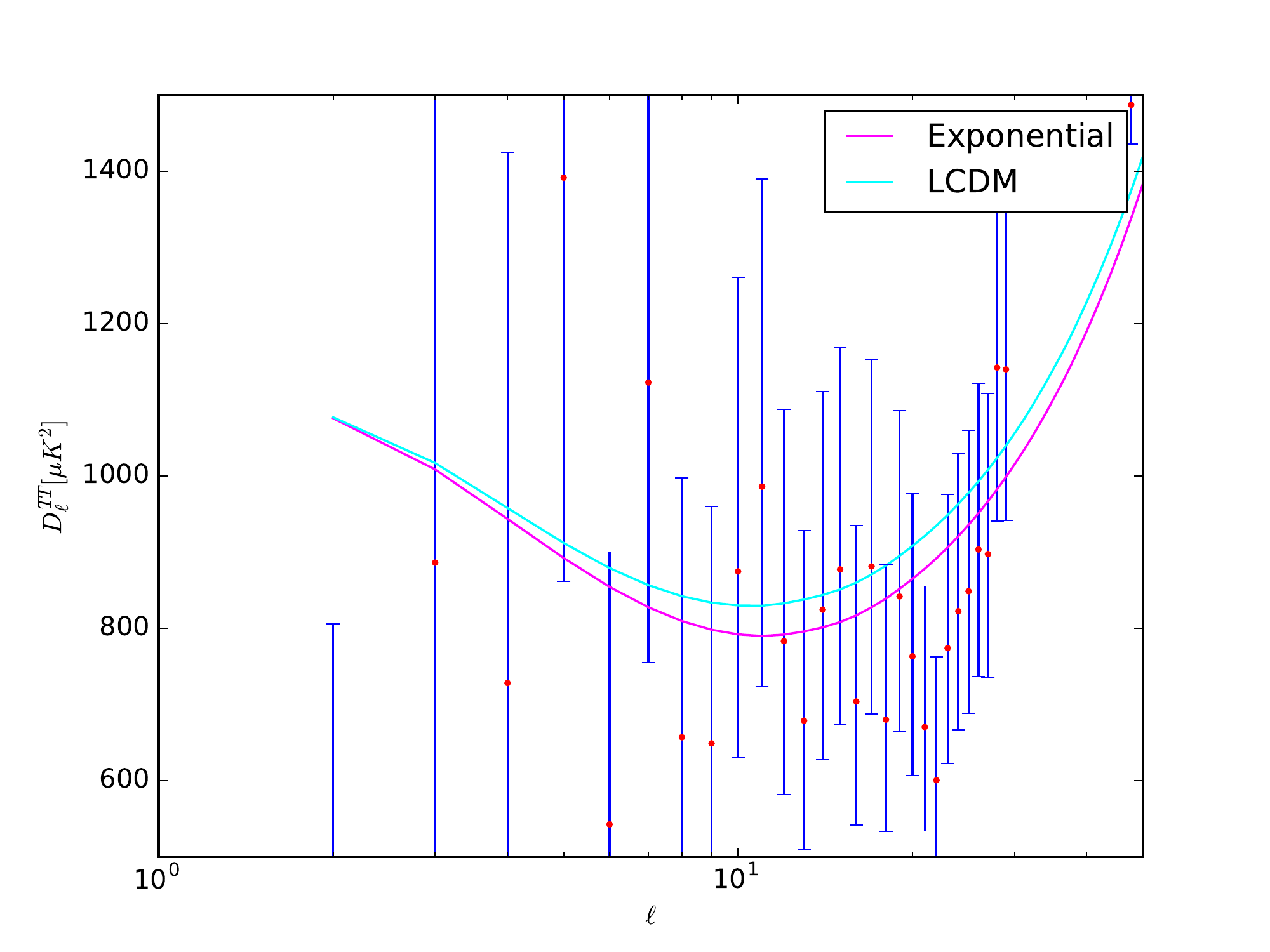}
\includegraphics[width=7.4cm]{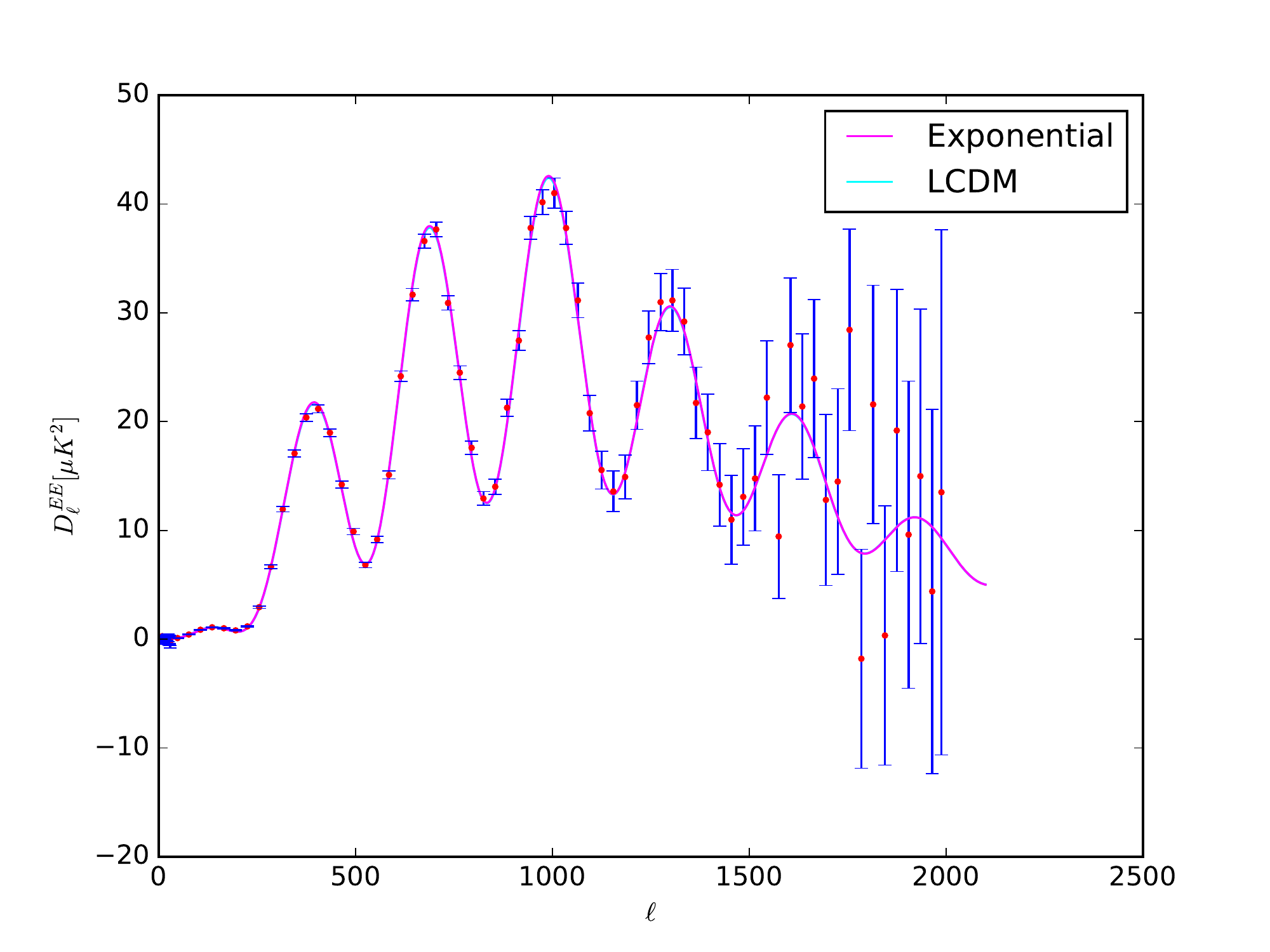}
\includegraphics[width=7.4cm]{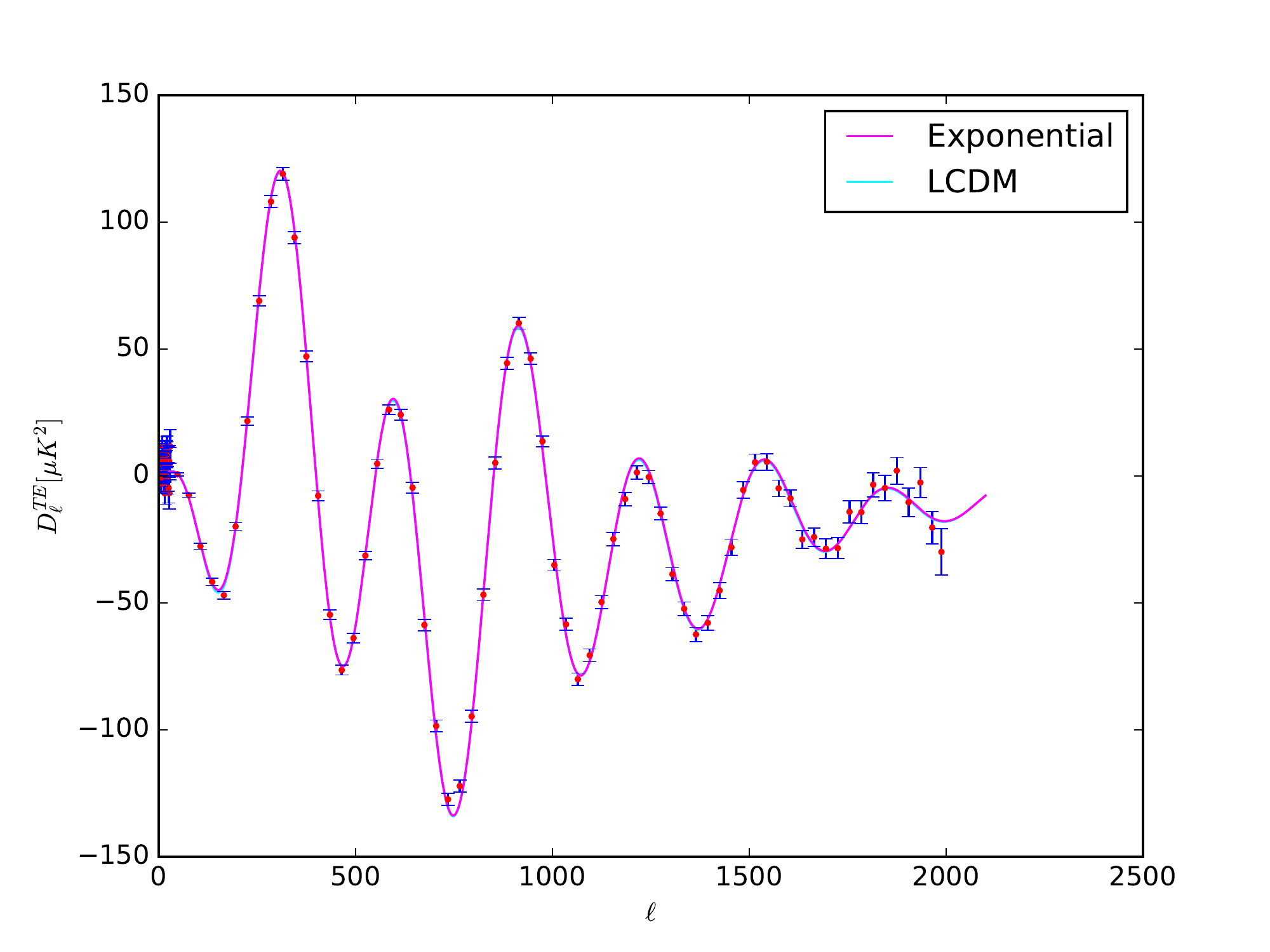}
\caption{Comparison of the temperature and polarization CMB angular power spectra computed for the best-fit of our modified Exponential model $\Lambda$CDM+r+$N_{\rm eff}$ (magenta) and the best-fit obtained with a minimal standard cosmological model $\Lambda$CDM+r (cyan), with Planck 2015 TT+lowTEB data (points with error bars). The oly difference between the two models is at lower-$\ell$ in the temperature power spectrum.}
\label{clbf}
\end{figure}

However, there is grounds for optimism: as we now show this result does not hold in a general $\Lambda CDM$ scenario expanded beyond its minimal 6 parameter model to allow for additional degrees of freedom. In fact even for the exponential model, which is completely ruled out in the 6 parameter minimal standard model, we find that, also with the modifications included, the modified exponential model becomes in full agreement with Planck and SN1a data simultaneously.

And this is our important finding: from Table~\ref{table2}, we notice that allowing for a single additional degree of freedom, in our case a free dark radiation component, then this model produces a slightly better fit of the data by gaining about a $\Delta \bar \chi^2=6.5$, when considering Planck TT+lowTEB, with respect to the $\Lambda$CDM+r model of the Table~\ref{table1}. All of this at the price of a single additional degree of freedom to extend to a 7 parameter $\Lambda CDM$.
So in this case varying the effective number of relativistic degrees of freedom, which are degenerate with the scalar spectral index $n_S$, allows the data to accommodate inflationary potentials that are otherwise ruled out, such as the 
counterexample of the modified exponential inflation we analyzed in this work.

\begin{figure}
\centering
\includegraphics[width=7.4cm]{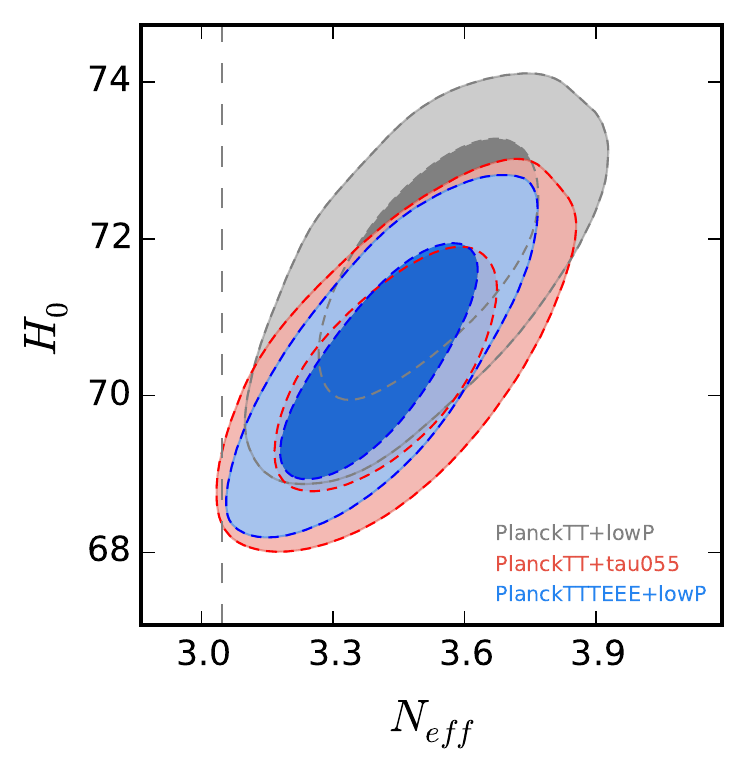}
\caption{Constraints at $68 \%$ and  $95 \%$ confidence levels on the $N_{\rm eff}$ vs $H_0$ plane, in our extended $\Lambda$CDM+r+$N_{\rm eff}$ scenario. In our modified Exponential scenario, we have a robust indication at more than $2\sigma$ for an extra dark radiation component, that allows the Planck data to be in agreement with the Hubble constant value found by Riess at al. in \cite{R16}.}
\label{fignnu}
\end{figure}

\begin{figure}
\centering
\includegraphics[width=7.4cm]{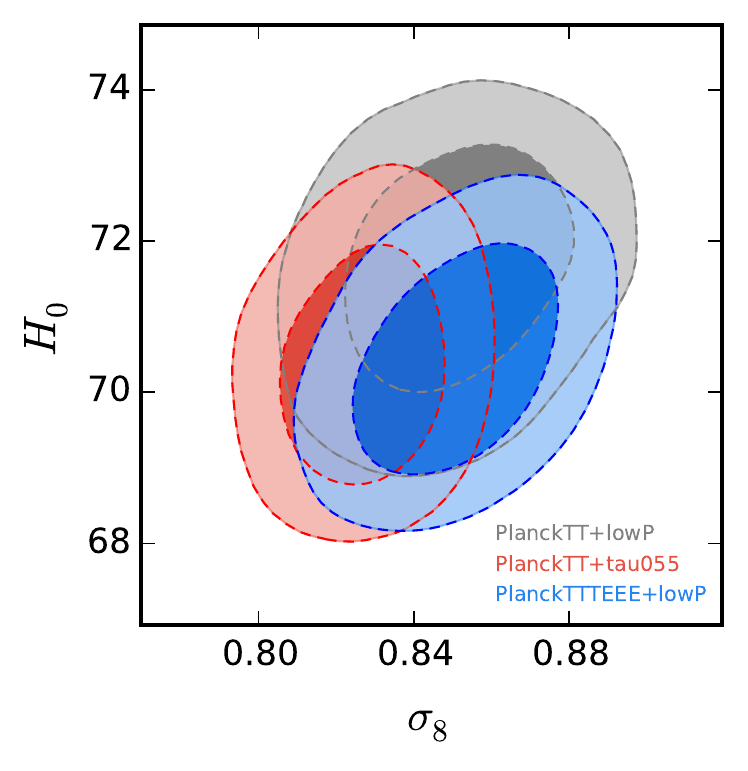}
\caption{Constraints at $68 \%$ and  $95 \%$ confidence levels on the $\sigma_8$ vs $H_0$ plane, in our extended $\Lambda$CDM+r+$N_{\rm eff}$ scenario. In our modified Exponential scenario, we don't have the strong degeneracy between them that is present without the entanglement corrections \cite{edvfrancois}.}
\label{figsigma8}
\end{figure}

Introducing a dark radiation component $N_{\rm eff}$ which is free to vary, to the modified exponential model investigated here, where the modification are derived from 
the theory of the origin of the universe from the quantum landscape multiverse, we find all the cosmological parameters with respect to $\Lambda$CDM+r model (see Table~\ref{table1}) shift.
In particular $\Omega_{\textrm{c}}h^2$  shifts toward higher values of more than 3$\sigma$, doubling its error bar.
As it has been previously shown in \cite{Tram:2016rcw,edvfrancois}, introducing a $N_{\rm eff}$ free to vary produces a value for this neutrino effective number higher than its expected value $3.045$ \cite{Mangano:2005cc,deSalas:2016ztq}, and a shift of all the parameters that are correlated with it. Also in our model of the modified exponential inflation, due to the strong correlation existing between $N_{\rm eff}$ and the Hubble constant $H_0$ (see Fig.~\ref{fignnu}), by increasing the neutrino effective number, we can relieve the tension between the constraints coming from the Planck satellite \cite{planckparams2013} and \cite{planckparams2015} and the local measurements of the Hubble constant of Riess at al. \cite{R11} and \cite{R16} \footnote{Other possibilities to relieve the tension has been considered by several authors (see for example \cite{DiValentino:2016hlg,Qing-Guo:2016ykt,edvlmh}), by introducing a dark energy equation of state $w<1$.}. For example, if we look at the results for Planck TT+lowTEB, we find $N_{\rm eff}=3.51\pm0.15$ and $H_0=71.6\, ^{+1.1}_{-0.9}$, with the error bars reduced by a half with respect to the minimal standard model. A $N_{\rm eff}>3.045$ means the presence of dark radiation, that can be explained by the existence of some extra relic component, as for example a sterile neutrino or a thermal axion \cite{darkradiation,edv1,ma1,edvfrancois, DiValentino:2015wba,DiValentino:2016ikp, Giusarma:2014zza, DiValentino:2013qma, Archidiacono:2011gq}. However, in contrast to the standard exponential scenario \cite{Tram:2016rcw,edvfrancois} the constraints for our model, where the introduced modifications are derived from 
the theory of the origin of the universe from the quantum landscape multiverse, are very robust. In our case the inclusion of the high-$\ell$ polarization data or the tau055 prior does not change the constraints in a significant way with respect to Planck TT+lowTEB (see Figs.~\ref{fignnu} and \ref{figsigma8}), even if the $\chi^2$ worsens as it happens without the modifications. Moreover, if we compute the value of $S_8=\sigma_8 \sqrt{\Omega_m/0.3}$, we find $S_8=0.831\pm0.023$ for Planck TT+lowTEB, which significanlty reduces 
the existing tension, that becomes at $1.9\sigma$, with KiDS-450 \cite{Hildebrandt:2016iqg} for which $S_8=0.745\pm0.039$, but this increases again at $2.3\sigma$  for Planck TTTEEE+lowTEB, for which we find $S_8=0.845\pm0.018$, as for the model without the modifications.

Since in our investigation we find a full consistency between the Planck data and the value of the Hubble constant measured in \cite{R16}, we can then safely add the prior $H_0=73.2\pm1.7$ km/s/Mpc, to check the stability of our results. This prior confirms the results found in the Planck TT+lowTEB and Planck TTTEEE+lowTEB cases, for which the $S_8$ tension with KiDS-450 is, respectively, at $1.8\sigma$ and $2.3\sigma$.

\begin{figure}
\centering
\includegraphics[width=7.4cm]{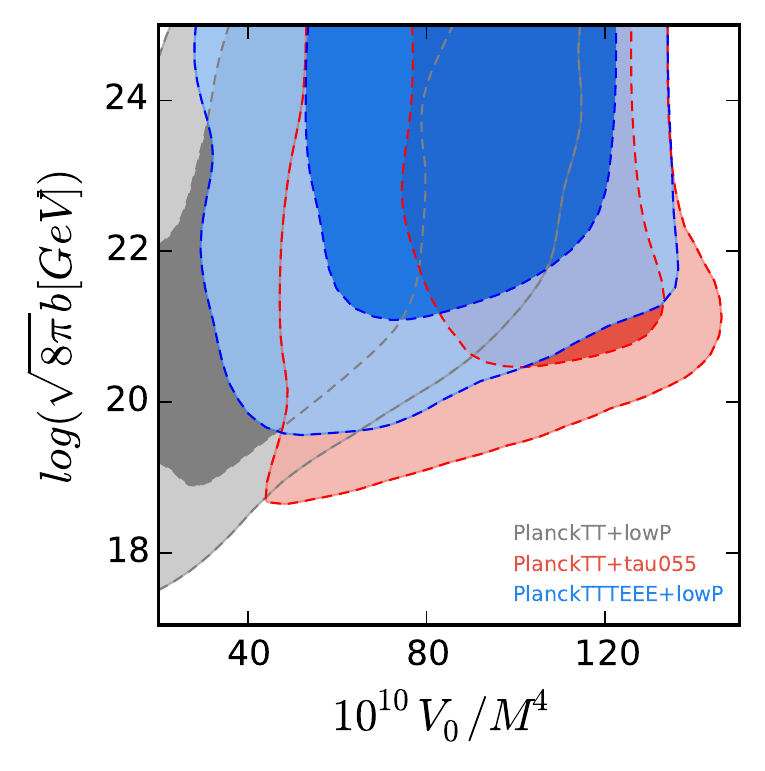}
\includegraphics[width=7.9cm]{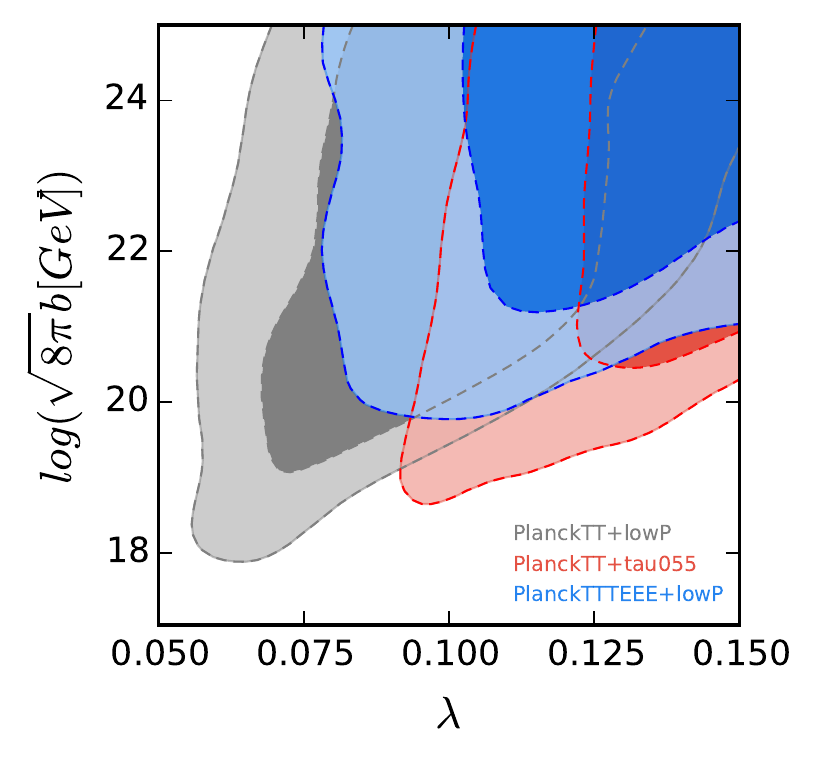}
\caption{Constraints at $68 \%$ and  $95 \%$ confidence levels on the $10^{10}V_0/M^4$ vs $log(\sqrt{8\pi}b)$ and $\lambda$ vs $log(\sqrt{8\pi}b)$ planes, in our extended $\Lambda$CDM+r+$N_{\rm eff}$ scenario.}
\label{figv0b}
\end{figure}

Regarding the inflationary parameters that describe the theory analyzed here, a constraint appears for the tensor-to-scalar ratio $r$, which is different from zero at more than $2$ standard deviations. We find for example, that $r=0.086\,_{-0.049}^{+0.024}$ for Planck TT+lowTEB. In Fig.~\ref{figv0b} there are shown the constraints at $68 \%$ and  $95 \%$ confidence levels in the $10^{10}V_0/M^4$ vs $log(\sqrt{8\pi}b)$ and $\lambda$ vs $log(\sqrt{8\pi}b)$ planes. We can see that the SUSY breaking scale is in between $2.6\times10^8<b<1.4\times10^{10}\, GeV$ and $V_0=5.7\, ^{+1.0}_{-3.0}\times 10^{-9} M_P^4$ for Planck TT+lowTEB; and we find $\lambda=0.098\,_{-0.026}^{+0.019}$.

\section{Conclusions}
\label{sec:conclusions}

If we insist in a 6 parameter standard model of cosmology, then the recent Planck datasets rule out convex potentials, including the exponential inflation we considered here as an illustration. This remains the case even in the presence of nonlocal modifications, like the one analyzed in this work, which is introduced by quantum entanglement in the theory of the origin of the universe from the quantum landscape. Meanwhile the friction between the findings of Planck \cite{planckparams2013,planckparams2015} and the recent SN1a measurements of the Hubble parameter of Riess et al \cite{R16} persists.

Our key finding is that if we allow for an extension beyond the 6 parameter standard model, by introducing dark radiation $N_{eff} > 3.045$, then the exponential inflationary potentials, modified by the theory analyzed here, produces a model that fits Planck datasets 2015 despite it being a convex potential, brings the Planck findings into agreement with the SN1a results of \cite{R16} for the Hubble parameter and naturally removes the friction between these datasets, without increasing the tension with weak lensing measurements, such as KiDS-450 \cite{Hildebrandt:2016iqg} and CFHTLenS \cite{Heymans:2012gg}, and, for which all the predictions made from this theory in \cite{tomolmh}  for the existence of CM anomalies, stand the scrutiny of Planck data \cite{planck2}.

\begin{figure}
\centering
\includegraphics[width=7.4cm]{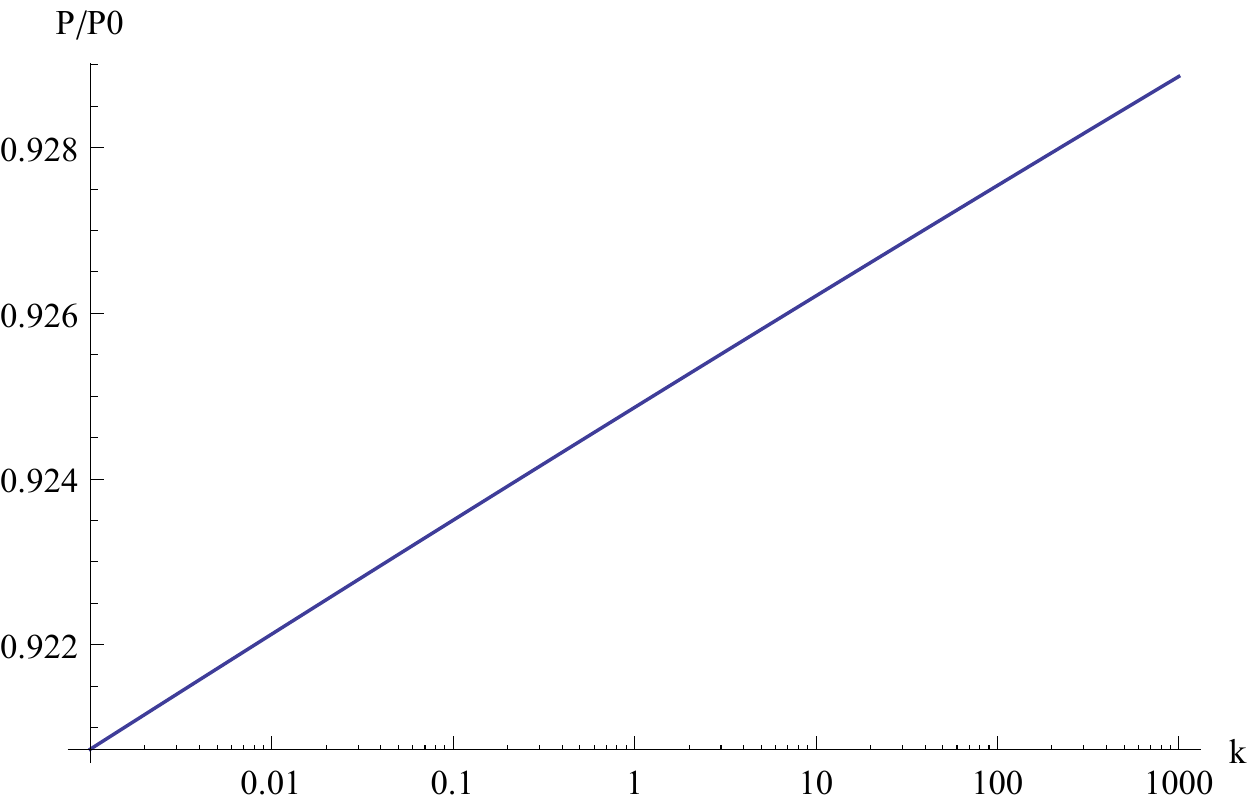}
\includegraphics[width=7.4cm]{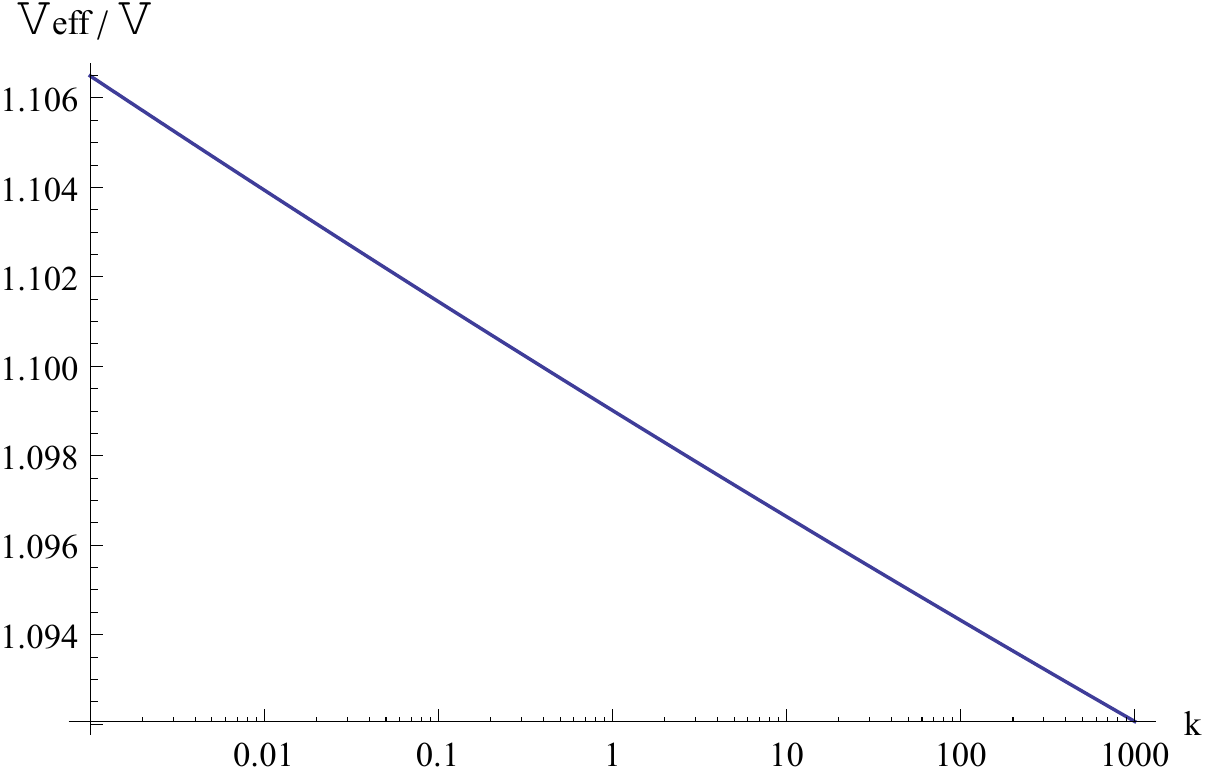}
\includegraphics[width=7.4cm]{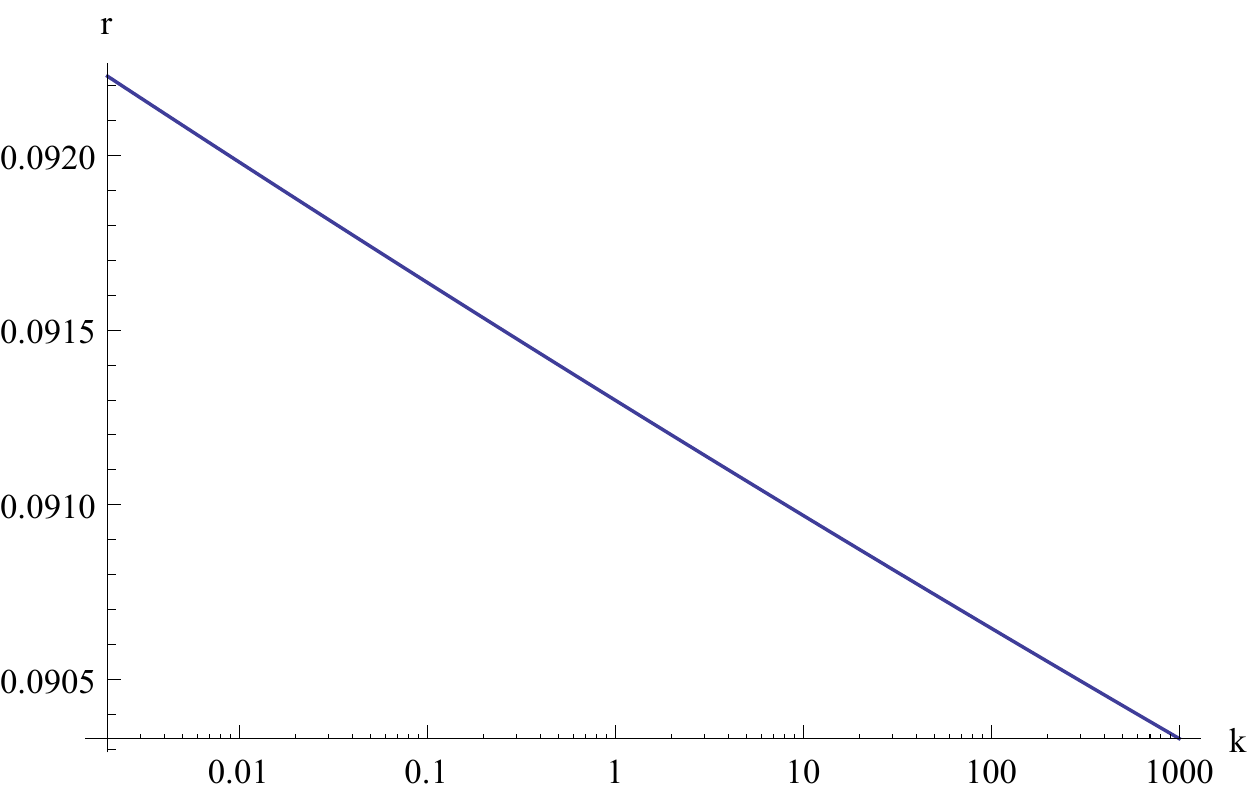}
\includegraphics[width=7.4cm]{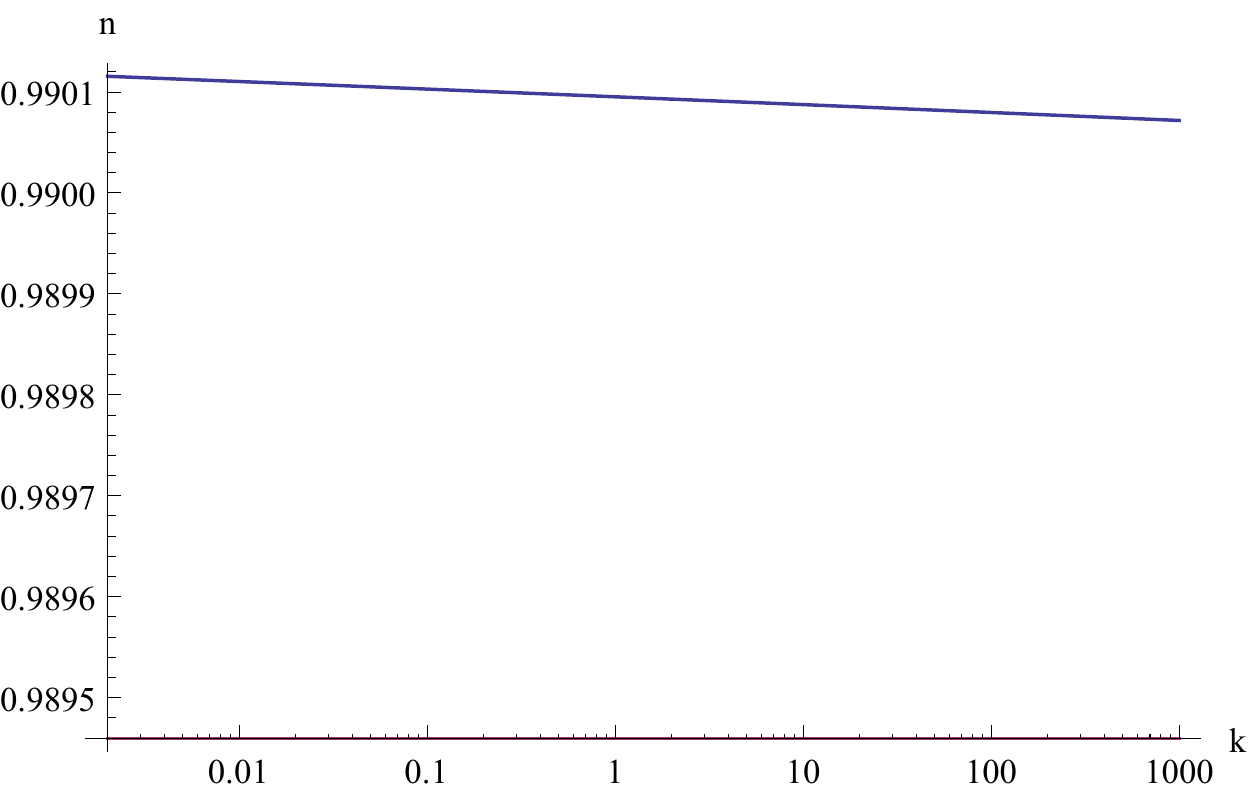}
\caption{Shown are: the ratio of the modified spectrum $P[k]$ over the unmodified $P0[k]$ spectra;the ratio of the effective potential $V_{eff}(\phi[k])$ versus the unmodified exponential potential $V(\phi_{0}[k])$; and the tensor to scalar ratio $r[k]$ and scalar index $n[k]$, for the $1-\sigma$ parameters of this model $b=7\times 10^{8} GeV , \lambda =0.098 , V_{0} = 5 \times 10^{-9}$ for the modified Exponential model.}
\label{cmbspectra}
\end{figure}

Our data analysis in the $\Lambda$CDM+r+$N_{\rm eff}$ scenario, constrains, for Planck TT+lowTEB dataset, the parameters $b$, $\lambda$ and $V_0$ to the following values for the $1\sigma$ and the $2\sigma$ allowed regions respectively: $2.6\times10^8<b<1.4\times10^{10}\, GeV$,  $\lambda=0.098\,_{-0.026}^{+0.019}$ and $V_0=5.7\, ^{+1.0}_{-3.0}\times 10^{-9} M_P^4$ at $68 \%$ c.l., and $b>4.8\times10^7\, GeV$,  $\lambda=0.098\,_{-0.038}^{+0.040}$ and $V_0<1.0\times 10^{-8} M_P^4$ at $95 \%$ c.l.. The strength of modification for the $1 \sigma$ value of parameters, shown in Fig.~\ref{cmbspectra}, is about $12\%$. As can be seen the modified scpetrum is more surpressed at lower multipoles than the higher ones.

The energy modification term $f(b, V)$ obtained from the analysis of the allowed parameters for the $1\sigma$ region, is up to $20 \%$ of the inflaton potential for these parameters. Therefore, the status of all the anomalies predicted in \cite{tomolmh} as tests of this theory, are in perfect agreement within the range observed by Planck 2015 \cite{planck2}. Furthermore, as mentioned when using these parameters to estimate the effective modified potential, we find that our model yields a robust higher value for the Hubble parameter, thus removing the friction between Planck and SN1a datasets, and our model makes a robust prediction for the existence of dark radiation $N_{eff} = 3.51\,\pm0.15$ at $68 \%$ c.l. which can be, for example, a light thermal axion in the form of dark matter or a sterile neutrino species.

A key question remains: is the agreement of the modified exponential model due to the presence of additional dark radiation or is it due to the nonlocal quantum entanglement modifications? A complete answer to this question would require a further comparison performed in terms of the Bayesian evidence. However, previously one of us analized the unmodified exponential inflation in the presence of the additional dark radiation in \cite{edvfrancois}, as previously done by \cite{Tram:2016rcw}, by considering several combination of datasets. The authors found in \cite{Tram:2016rcw} that with additional degrees of freedom the exponential model is in agreement with the Planck data and can solve the friction between the Planck and the SN1a \cite{R16} measurements of the Hubble parameter $H_0$. However, as it has been showed in \cite{edvfrancois} this extended model cannot solve the tension that is present between the Planck data and the weak lensing measurements of $S_8$, such as KiDS-450 \cite{Hildebrandt:2016iqg} and CFHTLenS \cite{Heymans:2012gg}, and when the tau055 prior is considered, the tension with $H_0$ is restored. Comparing that previous analysis of the unmodified exponential inflation with our very robust findings here, that do not change considering several combination of datasets, leads us to conclude that the agreement of the modified exponential inflation with all the data available, including the anomalies, the best fit region of inflationary parameters and the Planck and SN1a data on $H_0$ is due to the modifications. This fact makes this model very interesting, because in a very robust way, the model fits the variety of data and can resolve the friction among various observational findings.

 \acknowledgments
We would like to thank F. R. Bouchet and A. Melchiorri for stimulating discussions and we are grateful to the Planck Editorial Bord for taking the time to read the paper and approve it. This work has been done within the Labex ILP (reference ANR-10-LABX-63) part of the Idex SUPER, and received financial state aid managed by the Agence Nationale de la Recherche, as part of the programme Investissements d'avenir under the reference ANR-11-IDEX-0004-02. LMH acknowledges support from the Bahnson funds.

%\newpage

%%%%%%%%%%%%%%%%%%%%%%%%%%%%%%%%%%%%%%%%%%%%%%%%%%%%%%%%%%%%%%%%%%%%%%%%%%%%%%%
%\section*{References}
%%%%%%%%%%%%%%%%%%%%%%%%%%%%%%%%%%%%%%%%%%%%%%%%%%%%%%%%%%%%%%%%%%%%%%%%%%%%%%%
%\bibliography{References}

%\section{Additional Information}

% \begin{Competing Financial Interest.

\end{document}